# Generalized Measures for the Evaluation of Community Detection Methods

Vincent Labatut

Galatasaray University, Computer Science Department, Istanbul, Turkey
`vlabatut@gsu.edu.tr`
20/3/2013

**Abstract.** Community detection can be considered as a variant of cluster analysis applied to complex networks. For this reason, all existing studies have been using tools derived from this field when evaluating community detection algorithms. However, those are not completely relevant in the context of network analysis, because they ignore an essential part of the available information: the network structure. Therefore, they can lead to incorrect interpretations. In this article, we review these measures, and illustrate this limitation. We propose a modification to solve this problem, and apply it to the three most widespread measures: purity, Rand index and normalized mutual information (NMI). We then perform an experimental evaluation on artificially generated networks with realistic community structure. We assess the relevance of the modified measures by comparison with their traditional counterparts, and also relatively to the topological properties of the community structures. On these data, the modified NMI turns out to provide the most relevant results.

## 1 Introduction

Community detection is a part of the complex networks analysis field. It consists in characterizing the structure of such a network at the *mesoscopic* level, i.e. when considering neither the node (microscopic level) nor the whole network (macroscopic level), but rather an intermediary structure, called community. More concretely, one wants to break the network down to a set of loosely interconnected subgraphs, each one corresponding to a community. The problem is difficult to formalize, in the sense this task can be defined in many different ways. However, most authors agree on an intuitive description, which is to obtain communities whose nodes are more densely interconnected, compared to the rest of the network [1].

A document presenting a new community detection method generally has the following structure. First, the authors describe their algorithm in details. Second, they select some test data, which can be real-world and/or artificially generated networks, and apply both their algorithm and other existing tools to these data. Third, they process some measure to quantify the performances of the considered community detection tools. The resulting values are then used to compare these algorithms. The newly presented method generally happens to overcome the existing ones on one or several aspects (precision, speed, robustness, etc.).

This procedure raises several important methodological issues. When the test is performed on real-world data, according to which criteria should the networks be selected? Those networks display heterogeneous topological properties [2], some of which can introduce bias when comparing community detection methods. For example, a network with a low transitivity (a.k.a. clustering coefficient) will penalize clique-percolation methods looking for triangles. Moreover, in most case the actual community structure of real-world networks is not known with certainty: how reliable are the performance results obtained on such data? If the test is performed on some artificial data, then the selection of an appropriate generative model is a cause for concern. The results of the evaluation are supposed to be general enough to hold when the algorithms are applied to some real-world data. But for this to be true, the generative model must produce realistic networks, which is difficult to guarantee [3].

Despite the importance of these issues, in this article we put them apart to focus on another important methodological point: the tool used to measure the performance of the algorithms. In the literature, it *always* takes the form of a metric associating a numerical score to the community structure estimated by an algorithm for a given network. It is processed by comparing this estimated structure to the actual one, which is supposedly known for the considered network (either because it was identified, for a real-world network, or by construction for an artificial one). In the case of *mutually exclusive* communities, which interests us in this article, each community structure can be considered as a partition of the node set. Therefore, the standard approach to compare two community structures consists in quantifying the similarity between the two corresponding partitions. For this purpose, the most popular measures in the context of community detection are the Purity [4], the Adjusted Rand Index [5] and the Normalized Mutual Information [6].

However, as shown in a recent study [7], this approach has some limitations. Indeed, it is possible for two distinct community structures to be very close in terms of partition, therefore obtaining roughly the same score, and at the same time to have sensibly different topological properties (embeddedness, average distance, etc.). This, of course, is not desirable, since these properties should be considered to discriminate the community structures. One can trace back this problem to the cluster analysis origin of the measures. They completely ignore what makes the specificity of community detection: the structure of the network. In other words, the performance assessment is realized while ignoring a part of the available relevant information (the topological information).

In this article, we propose to modify certain existing measures in order to take the topological information into account. Our goal is to design a tool allowing a more relevant discrimination of the community structures. In the next section, we review the main measures used in the community detection literature to evaluate the performances of this type of algorithms. In section 3, we describe in details their limitation when applied to the comparison of community structures. We then propose our modifications in section 4, and evaluate them in section 5. We conclude with a discussion of our work and its possible extensions.

## 2    Traditional Approach

*Cluster analysis*, or unsupervised classification, is a part of the data mining field. It consists in partitioning a set of objects, in order to identify homogeneous groups. Each object is described individually through a vector of attributes, and the procedure is conducted by comparing objects thanks to these attributes. Community detection is obviously a very similar task, with one difference though. When considering complex networks, the objects of interest are nodes, and the information used to perform the partition is the network structure. In other words, instead of considering some individual information (attributes) like for cluster analysis, we take advantage of a relational one (links). However, the result is the same in both cases: a partition of the set of objects, which is called *community structure* in the context of complex network analysis.

It is therefore not surprising to see authors developing community detection tools use cluster analysis methods to assess the performance of their method. For some of them, the borrowing is explicit [8], whereas others developed their own tools, which happen to be similar to already existing ones [9, 10]. In cluster analysis, this assessment is performed thanks to a measure allowing to obtain a score representing the classifier performance. When a reference partition is available, this score represents the similarity between this actual partition and the one estimated by the considered classifier; and one refers to this measure as an *external evaluation criterion* [4]. A number of such measures exist, and in the domain of classification, the debate regarding which one is the most appropriate has been started a long time ago, and is still going on [11]: this shows how important this methodological point is. Indeed, what is the interest in evaluating a tool if the evaluation method is not valid?

A lot of the measures used in cluster analysis have been applied to community detection. However, three of them stand out in terms of popularity: Purity [4], Adjusted Rand Index [5] and Normalized Mutual Information [6]. Incidentally, each of them represents one of the three main families of measures designed as external evaluation criteria. In the first, each object is considered individually, whereas in the second the assessment is performed on pairs of objects. The third family relies on an information theory approach. For these two reasons, in this section we focus on these three measures.

In the rest of this article, we will note $X = \{x_1, \ldots, x_I\}$ and $Y = \{y_1, \ldots, y_J\}$ two partitions of the same set $S$, where $x_i$ and $y_j$ are the parts ($1 \leq i \leq I$ and $1 \leq j \leq J$). To denote the cardinalities, we use $n$ for the total number of elements in the partitioned set, and $n_{ij} = |x_i \cap y_j|$ for the intersection of two parts. We also note $n_{i+} = |x_i|$ and $n_{+j} = |y_j|$ the part sizes. When needed, elements will be represented by the variables $u$ and $v$.

## 2.1    Purity

The *Purity* measure [4] is historically the first one used in the context of community detection, since it was used by Girvan and Newman in their seminal article [9], under the name *fraction of correctly classified vertices*. More generally, the Purity appears in the literature under so many different names that it would be difficult to list them exhaustively.

The purity of a part $x_i$ relatively to the other partition $Y$ is expressed in the following way:

$$Pur(x_i, Y) = \max_j \frac{n_{ij}}{n_{i+}} \quad (1)$$

In other words, we first identify the part of $Y$ whose intersection with $x_i$ is the largest, and then calculate the proportion of elements in $x_i$ this intersection amounts to. The larger the intersection and the larger the purity, i.e. the larger the correspondence between the two considered parts. The total purity of partition $X$ relatively to partition $Y$ is obtained by summing the purity of each $x_i$, weighted by its prevalence in the considered set:

$$Pur(X, Y) = \sum_i \frac{n_{i+}}{n} Pur(x_i, Y) \quad (2)$$

The upper bound is 1, it corresponds to a perfect match between the partitions. The lower bound is 0 and indicates the opposite. It is important to notice the purity is not a symmetric measure: processing the purity of $Y$ relatively to $X$ amounts to considering the parts of $X$ majority in each part of $Y$. Therefore, in general, there is no reason to suppose $Pur(X, Y)$ and $Pur(Y, X)$ are equal.

From the community detection point of view, we can therefore use two distinct measures, depending on whether we calculate the purity of the estimated communities relatively to the actual ones, or the opposite. In cluster analysis, the first version is generally used, and called simply *Purity*, whereas the second version is the *Inverse Purity* [12]. In this document, we will use these terms to distinguish both versions.

It is difficult to determine which one of them was actually used in existing community detection works. Indeed, in their article, Girvan and Newman give a very succinct description of the measure they process [9]. A subsequent article seems to indicate it was the inverse purity [13] (note 19), which Newman directly confirmed to us. Many later works conducted by other authors used measures bearing the same name and/or directly referring to this article. However, due to the initial imprecision, it is very likely they used the purity in place of the inverse purity. For example, in [8] (p.4), Danon *et al.* make a comment on Girvan and Newman's measure, explaining how it can be biased by the number and sizes of communities. However, their remark is actually valid only for the purity, and not for the inverse version they are supposed to discuss.

This bias is an important limitation of both measures, and was also identified by the cluster analysis community. Purity tends to favor algorithms identifying numerous small communities. In the most extreme case, if the algorithm identifies $n$ communities containing a single node each, one gets a maximal purity, since each estimated community is perfectly pure. On the contrary, the inverse purity favors algorithms detecting few large communities. This time, the most extreme case occurs when the algorithm puts all the nodes in the same community. There again, one gets a maximal purity, because each actual community is perfectly pure: all the nodes it contains belong to the same (unique) estimated community. To solve this problem, Newman introduced an additional constraint [13]: when an estimated community is majority in several actual communities, all the concerned nodes are considered as misclassified.

The solution generally adopted in cluster analysis rather consists in processing the *F-Measure*, which is the harmonic mean of both versions of the purity [12]:

$$F(X,Y) = \frac{2 \cdot Pur(X,Y) \cdot Pur(Y,X)}{Pur(X,Y) + Pur(Y,X)} \quad (3)$$

The obtained measure is symmetric, and this combination is supposed to solve the previously mentioned bias. This approach penalizes in a similar way the under- and over-estimation of the number of communities. For this reason, we will later work with this adjustment, and not the one proposed by Newman.

### 2.2 Adjusted Rand Index

The *Rand Index* [14] is based on a different approach. Instead of directly considering how parts overlap, like the purity and other related measures, it focuses on pairwise agreement. For each possible pair of elements in the considered set, the Rand Index evaluates how similarly the two partitions treat them. One can distinguish 4 different cases. Let $a$ (resp. $d$) be the number of pairs in which nodes belong to the same part (resp. to different parts) in both partitions. Let $b$ (resp. $c$) be the number of pairs in which nodes belong to the same part in the first (resp. second) partition, whereas they belong to different parts in the second (resp. first) one. Formally, $a$ can be obtained by counting the number of pairs belonging to part intersections $x_i \cap y_j$:

$$a = \sum_{ij} \binom{n_{ij}}{2} \quad (4)$$

On the contrary, $b$ and $c$ correspond to pairs whose elements are located in different part intersections. For $b$, this amounts to counting the number of pairs belonging to part $x_i$ which were not already counted in $a$; and $c$ is defined symmetrically:

$$b = \sum_{i} \binom{n_{i+}}{2} - \sum_{ij} \binom{n_{ij}}{2} \quad (5)$$

$$c = \sum_{j} \binom{n_{+j}}{2} - \sum_{ij} \binom{n_{ij}}{2} \quad (6)$$

Finally, $d$ can be obtained by subtracting $a$, $b$ and $c$ to the total number of pairs. After simplification, we get:

$$d = \binom{n}{2} + \sum_{ij}\binom{n_{ij}}{2} - \sum_{i}\binom{n_{i+}}{2} - \sum_{j}\binom{n_{+j}}{2} \qquad (7)$$

Values $a$ and $d$ represent pairs for which both partitions agree, in the sense they both consider the nodes should be put together, or should be separated. On the contrary, $b$ and $c$ correspond to the two possible disagreements: in one partition the nodes are put together, whereas they belong to distinct parts in the other. The index is obtained by processing the proportion of pairs on which both partitions agree:

$$RI(X,Y) = \frac{a+d}{a+b+c+d} \qquad (8)$$

Like for the purity, the upper bound is 1, which corresponds to a perfect match between the partitions, and the lower bound is 0, which indicates the opposite. But unlike the purity, the Rand Index is symmetric: its value does not change if one switches the partitions.

In the domain of community detection, the chance-corrected version of this measure, called *Adjusted Rand Index* (ARI) [5], seems to be preferred. It is known to be less sensitive to the number of parts [15]. The chance correction is based on the general formula defined for any measure $M$ [16]:

$$M_{CC} = \frac{M - E(M)}{M_{max} - E(M)} \qquad (9)$$

Where $M_{CC}$ is the chance-corrected measure, $M_{max}$ is the maximal value $M$ can reach, and $E(M)$ is the value expected for some null model. Hubert & Arabie chose a model in which the partitions are generated randomly with the constraint of having fixed number of parts ($I$ and $J$) and part sizes ($n_{i+}$ and $n_{+j}$). Under this assumption, the expected value for the number of pairs in a part intersection $x_i \cap y_j$ is [5]:

$$E\left(\binom{n_{ij}}{2}\right) = \binom{n_{i+}}{2}\binom{n_{+j}}{2} / \binom{n}{2} \qquad (10)$$

Equation (10) can be used to process the expected values of $a$ and $d$, which in turn allows processing $E(RI)$. By replacing in equation (9) and after some simplifications, we get the final adjusted Rand index [5]:

$$ARI(X,Y) = \frac{\sum_{ij}\binom{n_{ij}}{2} - \sum_i\binom{n_{i+}}{2}\sum_j\binom{n_{+j}}{2}/\binom{n}{2}}{\frac{1}{2}\left(\sum_i\binom{n_{i+}}{2} + \sum_j\binom{n_{+j}}{2}\right) - \sum_i\binom{n_{i+}}{2}\sum_j\binom{n_{+j}}{2}/\binom{n}{2}} \qquad (11)$$

Like the Rand index, this measure is symmetric. Its upper bound is 1, meaning both partitions are exactly similar. Because it is chance-corrected, a value equal or below 0 represents the fact the similarity between $X$ and $Y$ is equal or less than what is expected from two random partitions.

### 2.3   Normalized Mutual Information

In cluster analysis, the use of the *Normalized Mutual Information* (NMI) is much more recent than for the previous measures [6]. It was introduced in the community detection domain by Danon *et al.* [8], and since then it has been used in many works. In this measures, both partitions $X$ and $Y$ are considered as discrete random variables, whose definition domains are $\{1,\ldots,I\}$ and $\{1,\ldots,J\}$, respectively. Their joint probability distribution is obtained by considering the frequencies measured on the available data:

$$p_{ij} = \frac{n_{ij}}{n} \quad (12)$$

The value $p_{ij}$ represents the probability, for a randomly drawn element, to belong simultaneously to parts $x_i$ and $y_j$. The marginal distributions are obtained by summing over the joint frequencies:

$$p_{i+} = \sum_j p_{ij} \quad (13)$$

$$p_{+j} = \sum_i p_{ij} \quad (14)$$

The value $p_{i+}$ (resp. $p_{+j}$) represents the probability, for a randomly drawn element, to belong to part $x_i$ (resp. $y_j$). From there, one can process the mutual information $I(X,Y)$ of these variables, which measures the probabilistic dependence between them [17]:

$$I(X,Y) = \sum_{i,j} p_{ij} \log \frac{p_{ij}}{p_{i+} p_{+j}} \quad (15)$$

The mutual information corresponds to the quantity of information shared by the variables. Unlike the purity, but like the Rand Index, it is symmetric. Its lower bound is 0, representing the independence of the variables (they share no information). The upper bound corresponds to a complete redundancy, however this value is not fixed. Several normalizations exist to solve this problem. The approach used in [18], and later by Danon *et al.* and the rest of the community detection field, consists in dividing the mutual information by the arithmetic mean of the entropies: $NMI(X,Y) = 2I(X,Y)/\big(H(X) + H(Y)\big)$, where $H(X) = -\sum_i p_{i+} \log(p_{i+})$ and $H(Y) = -\sum_j p_{+j} \log(p_{+j})$. The final expression of this measure is therefore:

$$NMI(X,Y) = \frac{-2\sum_{ij} p_{ij} \log(p_{ij}/p_{i+} p_{+j})}{\sum_i p_{i+} \log(p_{i+}) + \sum_j p_{+j} \log(p_{+j})} \quad (16)$$

This normalization retains the lower bound and symmetry of the measure, and its upper bound becomes 1.

## 3     Limitations of the Existing Measures

By definition, all the measures coming from cluster analysis, including the three presented in the previous section, consider a community structure only as a partition of the node set. In the context of community detection, this can be viewed as a limitation, because all classification errors do not necessarily have the same importance.

Let us consider the example presented in Fig. 1, which displays a network containing two communities, each one represented by a different color. This community structure is noted $R$, and the red (i.e. left) and blue (i.e. right) communities are noted $r_1$ and $r_2$, respectively. We propose two different estimations $A$ and $B$ of this reference community structure. For both of them, the left and right communities are numbered 1 and 2, respectively. Each one of these estimated community structures includes a classification error: one node from the left community is incorrectly placed in the right one. For $A$, this misclassification concerns node 2, whereas for $B$ it is node 6.

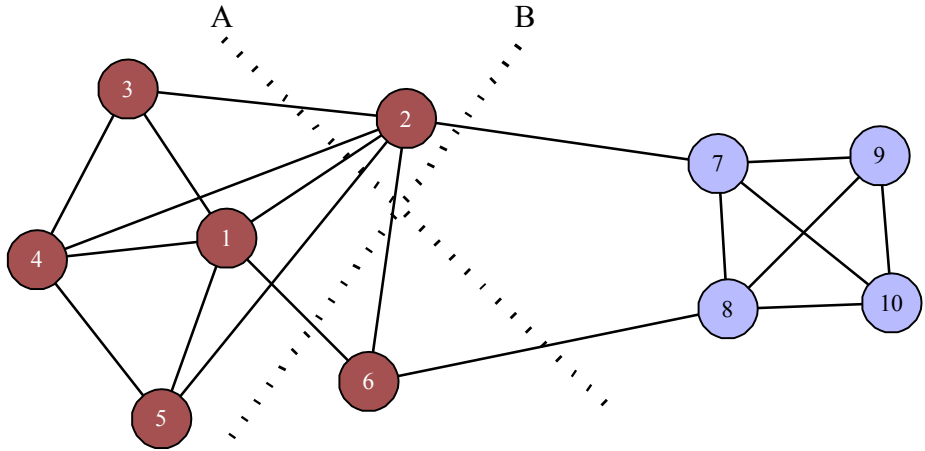

**Fig. 1.** Example illustrating the limitation of purely partition-based measures. Colors correspond to the actual communities, whereas lines labeled $A$ and $B$ represent two different (incorrect) estimations of this community structure.

Let us apply the measures presented in the previous section, in order to compare $A$ to $R$. We obtain the score 0.9 for both the purity and inverse purity. Consequently, the F-Measure, which is their mean, reaches the same value. For the adjusted Rand index, we get 0.6, and 0.62 for the NMI. Now, if we process the same measures for the other partition $B$, we get exactly the same values. Indeed, from a partition perspective, nothing allows to distinguish nodes 2 from node 6, so misclassifying the former or the latter leads to the same evaluation. In other words, those measures consider the errors present in $A$ and $B$ to be exactly similar.

Yet, intuitively, those errors do not seem equivalent at all. Indeed, node 2 is much more integrated in its actual community than node 6. Its misclassification in partition *A* is therefore a more serious error than that of node 6 in partition *B*. In consequence, the score associated to *B* should be higher.

To check more objectively this intuition, we can consider the modularity [19] of these partitions. This measure quantifies the quality of a community structure in a blind way, i.e. without the use of a reference. In cluster analysis terms, it would be called an internal evaluation criterion (cf. the introduction of section 2). For this matter, it compares the proportion of links located inside the communities with the expectation of the same value for a random model generating similar networks (same size and degree distribution). The modularity has been used as an objective function by numerous community detection algorithms [1]. In our case, the reference *R* reaches a modularity of 0.36, whereas *A* and *B* obtain the scores 0.18 and 0.34, respectively. More than their magnitude, what is relevant here is the relative differences between those values: *A* clearly leads to a lower score than *B*, which confirms our intuition.

This observation, performed on our very simple example, is corroborated on more realistic networks by the recent study by Orman *et al*. [7]. Its authors compare community structures by considering traditional measures (such as those presented in the previous section), but also the distribution of several measures allowing to characterize them topologically (community size, transitivity, density, etc.). One of their conclusions is that two community structures can at the same time reach very similar scores, and be topologically very different. There are two important consequences to this result. First, an estimated community structure can reach a high score, without necessarily being topologically similar to the actual community structure. Second, two estimated partitions can reach approximately the same score without having automatically the same topological properties. Because of these limitations, we can state traditional measures are not perfectly adapted neither to the evaluation of a community detection algorithm in absolute terms, nor to the comparison of several such algorithms.

## 4    Proposed Modifications

Of course, the problem highlighted in the previous section comes from the fact the traditional measures consider a community structure is simply a partition, and therefore ignore a part of the available information: the network topology. In order to make a more reliable evaluation, Orman *et al*. propose to jointly use traditional measures and various topological properties [7]. However, they also acknowledge this makes the evaluation process more complicated, due to the multiplicity of values to take into account.

The solution we propose here, on the contrary, consists in retaining a single value, by modifying traditional measures so that they take the network topology into account. This approach allows benefiting from the concision of a unique score to measure and compare community detection algorithms.

In this section, each one of the first three subsections presents the proposed modifications for one of the three measures described in section 2. All of those modifications are based on the definition of an individual weight, reflecting the relative importance of each node. We chose to discuss it separately, in the last subsection, because the nature of this weight constitutes a separate point, independent from the general form of the modified measures.

### 4.1   Modified Purity

Compared to the measures of the two other families, the purity has the advantage that it can be expressed in order to make appear the individual contribution of each node to the total score. For this purpose, we first define the notion of purity of a node $u$ for a partition $X$ relatively to another partition $Y$:

$$Pur(u, X, Y) = \delta\left(\arg\max_j n_{\alpha j}, \beta\right) \qquad (17)$$

Where $u \in x_\alpha$ and $u \in y_\beta$ ; and $\delta$ is the Kronecker delta, i.e. $\delta(a, b) = 1$ if $a = b$, and 0 otherwise. The function is therefore binary: 1 if the part of $Y$ containing $u$ is majority in that of $X$ also containing $u$, and 0 otherwise. As an example, consider $Pur(2, A, R)$ in the case of Fig. 1. In $R$, node 2 belongs to the red (left) part $r_1$, so the second argument of the $\delta$ is 1. In $A$, it belongs to the right part, whose intersection is larger with $r_2$ than with $r_1$, so the first argument of the $\delta$ is 2. Consequently, $Pur(2, A, R) = \delta(2,1) = 0$. On the contrary, if we focus on node 6 instead, we obtain $Pur(6, A, R) = \delta(1,1) = 1$.

The purity of a part $x_i$ relatively to a partition $Y$ can then be calculated by averaging the purity of its nodes:

$$Pur(x_i, Y) = \frac{1}{n_{i+}} \sum_{u \in x_i} Pur(u, X, Y) \qquad (18)$$

The above expression is equivalent to that of equation (1), thus it allows deriving the total purity of partition $X$ relatively to $Y$, as in equation (2). By developing the resulting expression, we get:

$$Pur(X, Y) = \sum_i \sum_{u \in x_i} \frac{1}{n} Pur(u, X, Y) \qquad (19)$$

One can notice the purity of each node is weighted by a value $1/n$. In order to take into account the topological information, we propose to replace this uniform weight by a value $w_u$, which can be different for any node $u$. Its role is to penalize more strongly misclassifications concerning topologically important nodes. We then get the modified purity $Pur'$, defined as follows:

$$Pur'(X, Y) = \sum_i \sum_{u \in x_i} \frac{w_u}{w_+} Pur(u, X, Y) \qquad (20)$$

Where $w_+ = \sum_v w_v$, i.e. the sum of all weights. This normalization allows keeping the measure between 0 et 1. Finally, by applying to $Pur'$ the same principle we described in equation (3) (i.e. taking the harmonic mean of the purity and inverse purity), we obtain a modified F-Measure, which takes the network topology into account, and that we note $F'$.

### 4.2 Modified ARI

Because the Rand index is based on pairwise comparisons, it is not possible to isolate the individual effect of each node, like we did for the purity. However, we can proceed similarly for pairs of nodes. In the original measure, each pair contributes similarly to the total score. Instead, we propose to distinguish them in terms of topological importance.

The most direct approach consists in associating a specific weight to each pair of nodes. For instance, one could consider the geodesic distance between the nodes. The consequence would be to penalize more disagreements on pairs of distant nodes. However, there is no reason to think misclassifications on distant nodes are more important than on close ones (or the opposite).

Using nodal weights like for the purity seems to be a more appropriate solution. Since we handle pairs of nodes $(u, v)$ here, we propose to use the product of the two corresponding nodal weights: $w_u w_v$. Of course, any other combination could be used, but our goal was to clearly advantage couples of important nodes. Then for any subset $s$ of $S$, we define the following quantity:

$$W(s) = \sum_{u,v \in s} w_u w_v \qquad (21)$$

The binomial coefficients used in the formulas of the original and adjusted Rand indices aim at counting the number of pairs present in various subsets of the partitions. This amounts to processing $W$ in the specific case where all $w$ are equal to 1. In order to obtain the modified versions of these measures, we simply replace all binomial coefficients by our generalized coefficient $W$, in their respective formulas. Therefore, from equation (11) we get the modified version of the ARI, noted $ARI'$:

$$ARI'(X,Y) = \frac{\sum_{ij} W(x_i \cap y_j) - \sum_i W(x_i) \sum_j W(y_j)/W(S)}{\frac{1}{2}\left(\sum_i W(x_i) + \sum_j W(y_j)\right) - \sum_i W(x_i) \sum_j W(y_j)/W(S)} \qquad (22)$$

### 4.3 Modified NMI

In the traditional definition of the NMI, one implicitly considers all nodes have the same probability $1/n$ to be randomly drawn. This becomes explicit if we rewrite the expression of $p_{ij}$ given in equation (12) in the following way:

$$p_{ij} = \sum_{u \in x_i \cap y_j} \frac{1}{n} \qquad (23)$$

We propose to replace this uniform value by the node-specific weight $w$ already introduced for the previous measures. As before, it must be normalized using $w_+$, in order to sum to 1. We can consequently define a modified joint probability distribution $p'_{ij}$:

$$p'_{ij} = \sum_{u \in x_i \cap y_j} \frac{w_u}{w_+} \qquad (24)$$

By replacing $p_{ij}$ by $p'_{ij}$ in equations (13) and (14), we obtain $p'_{i+}$ and $p'_{+j}$, respectively. We then use these modified probability distributions in the definition given in equation (16), in order to get the modified normalized mutual information, noted $NMI'$:

$$NMI'(X,Y) = \frac{-2\sum_{ij} p'_{ij} \log(p'_{ij}/p'_{i+}p'_{+j})}{\sum_i p'_{i+} \log(p'_{i+}) + \sum_j p'_{+j} \log(p'_{+j})} \qquad (25)$$

### 4.4    Nodal Weights

All the modified measures we proposed in this section depend on the definition of an individual weight $w_u$ representing the relative importance of each node $u$ in the considered network. The question is therefore now to determine how to characterize and quantify this importance. Our general idea is that a misclassification concerning a node strongly integrated into its community should count more than for a node located on the community fringe. For this purpose, we can consider the node degree. This way, we give more weight to community hubs such as node 2 from Fig. 1, and less weight to peripheral nodes such as node 6. In order to get a normalized value, we divide by the maximal degree observed in the network, leading to the *normalized degree* $d'$:

$$d'(u) = \frac{d(u)}{\max_v d(v)} \qquad (26)$$

Where $d(u)$ denotes the degree of node $u$. This value ranges from 0 (no connection at all) to 1 (most connected node in the whole network).

However, this approach can be criticized on two points. First, it is possible for a high degree node to have its connections distributed over numerous communities, therefore preventing any strong integration into any particular community. Since the community membership of this node seems rather uncertain, giving it a large weight appears inappropriate. Second, using only the degree leads to downplaying the importance of nodes whose connections are few, but entirely located inside their community. The *embeddedness measure e* [2] allows solving both problems:

$$e(u) = \frac{d_{int}(u)}{d(u)} \qquad (27)$$

Where $d_{int}(u)$ is the internal degree of node $u$, i.e. the number of connections it has in its own community. Thus, the embeddedness is the proportion of neighbors located in the same community than the node of interest. It ranges from 0 (no neighbor in the same community) to 1 (all neighbors in the same community).

In order to combine the normalized degree and embeddedness, we propose to multiply them. This way, the more a node possesses both properties and the more it is important for us. The weight is therefore $w_u = d'(u)e(u)$, which after simplification leads to the following expression:

$$w_u = \frac{d_{int}(u)}{\max_v d(v)} \qquad (28)$$

Note we treated the question of the nodal weight independently from the measure modifications for two reasons. First, this point is common to all three modifications we proposed, in the sense each of them needs this weight. Second, the specific weight described above is only a proposal: it can be adapted depending on the user's needs. For instance, if the links of the considered network are weighted, one can consider the *strength* of the nodes instead of their degree.

By using a uniform $w_u$ for every node, we obtain the traditional version of the considered measure. Thus, the modifications we propose can be considered as generalization of the traditional measures.

**Table 1.** Values obtained for the community structures displayed in Fig. 1.

| Measure | Traditional | Modified | |
|---|---|---|---|
| Partition | *A* and *B* | *A* | *B* |
| F-Measure (Purity) | 0.90 | 0.85 | 0.94 |
| Adjusted Rand Index | 0.60 | 0.44 | 0.75 |
| Normalized Mutual Information | 0.62 | 0.52 | 0.72 |

Let us now consider again the example from Fig. 1, and process the modified measures for both estimated partitions. Table 1 recapitulates the previous and newly calculated values. For all three measures, the scores obtained for partition *A* are lower than those of partition *B*, which is the behavior we were expecting, as explained in section 3.

## 5  Experimental Evaluation

The results obtained on the example from Fig. 1 are obviously not sufficient to assess the relevance of the modified measures. We therefore applied them on a larger dataset. In this section, we first present the experimental setup we used. Then, we describe how the proposed modification affected the individual performance scores. Finally, we study its effect on algorithm ranking.

### 5.1  Setup

We decided to use the same data, and to apply the same community algorithms than in [7] for our experimental validation, for several reasons. First, this study by Orman *et al.* contains observations regarding the topological properties of both real and estimated community structures. They used them to illustrate how community structures obtaining similar traditional scores can in fact be sensibly different, topologically. Thanks to them, we will be able to verify if our modified measures behave as expected, i.e. are sensitive to these differences. Moreover, they used artificially generated networks, which means the real community structures are known with certainty. Finally, the generative model they selected reaches the highest possible level of realism, at least according to current knowledge on real-world systems. This point is important, in order to be able to generalize our results.

The dataset is constituted of 5 networks of 25000 nodes each, whose main topological properties are consistent with real-world networks studied in the literature: degree distribution, transitivity (clustering coefficient), community sizes, embeddedness, etc. Eight different community detection algorithms are applied to these networks, in order to estimate the community structures. They are recent and representative enough of the main methods designed to perform community detection: Copra [20], FastGreedy [13], InfoMap [21], InfoMod [22], Louvain [23], MarkovCluster [24], Oslom [25] and WalkTrap [26]. Since those topics are not the main point of this article, we refer the reader to [7] for any further details regarding the generative process and community detection algorithms. We also insist on the fact our goal with this work is not to identify the best algorithm (which, as mentioned in the introduction, necessitates tackling a number of methodological problems), but rather to check the relevance of the evaluation tool we propose (i.e. the modified measures).

In their work, Orman *et al*. use a representative set of traditional measures to compare the partitions estimated by the considered algorithms: the fraction of correctly classified nodes (i.e. Newman's purity, as explained in section 2), the Rand index and its adjusted version, and the NMI. For the adjusted Rand index and NMU, we can directly use their results and compare them with those obtained for the corresponding modified versions described in section 4. However, it is not possible to do so for the purity, since we need to compare our modified measure to the F-Measure in order to make a relevant evaluation. Therefore, we had to compute the F-Measure ourselves.

### 5.2    Effect on the Scores

Fig. 2 displays the results obtained for all the considered measures. The values for the traditional versions are on the left side of the plot, whereas those for the modified ones are on the right. For each measure, the algorithms are ordered by decreasing value of the traditional version. In order to ease visual comparison, the same order is kept for the modified version. This allows highlighting disagreements between both versions.

Globally, the performances increase for all measures when comparing the traditional and modified versions. We used Student's $t$ test ($\alpha = 0.05$) to assess the significance of this evolution. InfoMod is the only algorithm to undergo a decrease with all three measures, and moreover those are significant. WalkTrap increases or decreases depending on the measure, but never significantly.

The algorithm with the largest improvement is by far Louvain: highest for the F-Measure (+0.2) and ARI (+0.25), second highest for the NMI (+0.07). The typical improvement is rather under 0.1 for the other algorithms. At a lesser extent, FastGreedy also experiences a significant performance improvement for all three measures. Oslom and MarkovCluster see their performance significantly increase, but only for the F-Measure and ARI. For the NMI, we observe a decrease and an increase, respectively, but those are not significant.

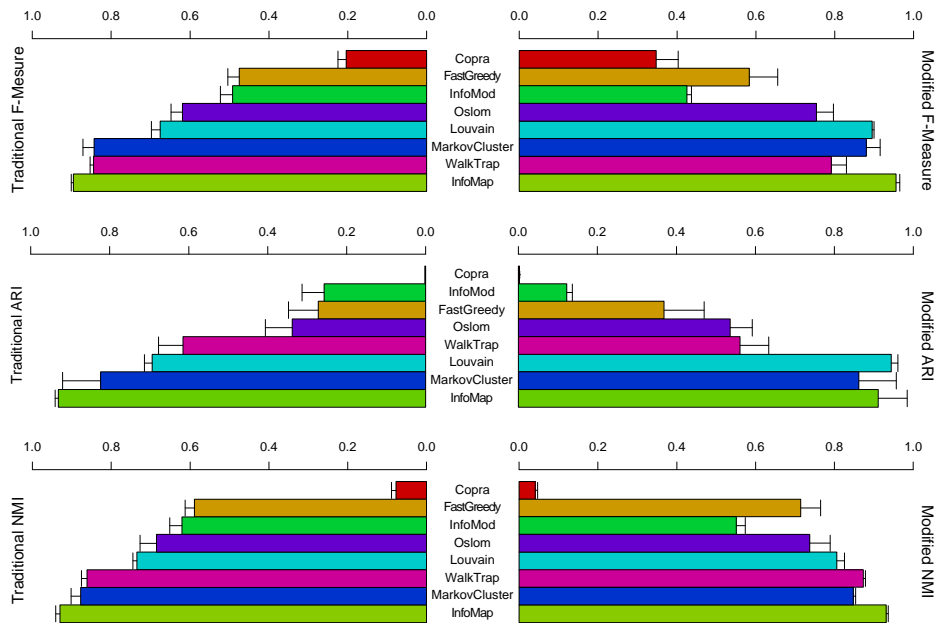

**Fig. 2.** Comparison of the results obtained with the traditional (left) and modified (right) versions of the measures.

InfoMap undergoes a slight improvement with the F-Measure, whereas for the ARI and NMI it decreases slightly, but not significantly. This might be due to the fact its performance is already so high with the traditional versions that there is not much room for increase.

Finally, the measures do not agree for Copra, which undergoes a clear increase with the F-Measure (+0.18), a small decrease (−0.04) with the NMI and a non-significant change with the ARI.

By construction of the modified measures, an increase in the score of a given algorithm can be interpreted as the fact its performance relies mainly on nodes with high weights, i.e. of larger topological importance. Therefore, according to the three considered measures, this is the case of most of the algorithms, except Copra and InfoMod (the latter of which they all agree upon). Interestingly, these are also ranked last by all measures, be it the traditional of modified versions. This means that, amongst the considered algorithms, those obtaining the lowest performance from a purely partition-based perspective are also those who do not seem to be good on topologically important nodes.

### 5.3   Effect on the Rankings

Table 2, Table 3 and Table 4 display the algorithms ranked by performance according to both versions of the ARI, F-Measure and NMI, respectively. To order them, we performed an ANOVA and applied Tukey's test with a significance level of $\alpha = 0.05$. Algorithms whose scores are not considered significantly different were placed on the same row. In our analysis, we focus on the correspondences and discrepancies identified by Orman *et al*. between the traditional measures and the topological properties. We discuss if the modified versions of the measures allow consistently taking this aspect of the performance into account.

**Table 2.** Algorithm rankings obtained with both traditional and modified versions of the ARI. Algorithms experiencing a change in their relative position are represented in bold.

| Traditional Adjusted Rand Index | | Modified Adjusted Rand Index | |
| --- | --- | --- | --- |
| Rank | Algorithm | Rank | Algorithm |
| 1 | InfoMap, MarkovCluster | 1 | **Louvain**, InfoMap, MarkovCluster |
| 3 | Louvain, WalkTrap | 4 | WalkTrap, **Oslom** |
| 5 | Oslom, FastGreedy, InfoMod | 6 | FastGreedy |
| 8 | Copra | 7 | **InfoMod** |
| - | - | 8 | Copra |

It is worth noticing the traditional versions of the F-Measure and NMI give exactly the same rankings. For this reason, we will discuss them jointly. But first, we start with the ARI. According to its traditional version, there is no significant difference between InfoMap and MarkovCluster. However, the topology-based observations show the former is much closer to the reference structure. For this reason, we would expect the modified version to make a distinction between them. However, this is not the case: no significant difference is detected.

The traditional version does not make any significant distinction between Louvain and WalkTrap. But topologically speaking, WalkTrap is supposed to be the closest to the reference just after InfoMap, so we would expect this difference to appear in the ranking based on the modified version. Nevertheless, we observe the opposite: Louvain is inconsistently raised to the level of InfoMap and MarkovCluster.

Oslom, FastGreedy and InfoMod are considered to have equivalent performance by the traditional version. From a topological point of view, Oslom and FastGreedy are very close to Louvain, this one being slightly better. As mentioned before, Louvain has indeed a better rank according to the modified version. However, our measure also introduces a distinction between FastGreedy and Oslom. Concerning InfoMod, it is supposed to be much topologically different from the reference than Oslom and FastGreedy. This is consistently reflected with the modified measure.

**Table 3.** Algorithm rankings obtained with both traditional and modified versions of the F-Measure. Algorithms experiencing a change in their relative position are represented in bold.

| Traditional F-Measure | | Modified F-Measure | |
|---|---|---|---|
| **Rank** | **Algorithm** | **Rank** | **Algorithm** |
| 1 | InfoMap | 1 | InfoMap |
| 2 | MarkovCluster, WalkTrap | 2 | **Louvain**, MarkovCluster |
| 4 | Louvain, Oslom | 4 | **WalkTrap**, Oslom |
| 6 | InfoMod, FastGreedy | 6 | FastGreedy |
| 8 | Copra | 7 | **InfoMod**, Copra |

We now turn to the F-Measure and NMI. For both of their traditional versions, InfoMap is ranked first, and alone. This is also the case with the modified versions, which is consistent with the topology.

According to the traditional versions, WalkTrap and MarkovCluster perform equivalently. Both modified measures manage to make a distinction between them, but they disagree. For the F-Measure, MarkovCluster is better, which is inconsistent with our topological knowledge, whereas on the contrary the NMI consistently puts WalkTrap at a higher rank.

Louvain and Oslom obtain the same rank with the traditional versions. From a topological point of view, we know they are indeed very close, the former being slightly closer to the reference. The modified F-Measure makes the correct distinction between them, but tends to overestimate their ranking, putting Louvain at the level of MarkovCluster and Oslom at that of WalkTrap. The modified NMI keeps on considering the algorithms are not significantly different, which seems more relevant.

The traditional versions consider InfoMod and FastGreedy have similar performance. For both modified versions, InfoMod is ranked lower, which is consistent with the topology-based observations. The NMI additionally raises FastGreedy to the level of Louvain and Oslom, which is consistent.

**Table 4.** Algorithm rankings obtained with both traditional and modified versions of the NMI. Algorithms experiencing a change in their relative position are represented in bold.

| Traditional NMI | | Modified NMI | |
|---|---|---|---|
| **Rank** | **Algorithm** | **Rank** | **Algorithm** |
| 1 | InfoMap | 1 | InfoMap |
| 2 | MarkovCluster, WalkTrap | 2 | WalkTrap |
| 4 | Louvain, Oslom | 3 | **MarkovCluster** |
| 6 | InfoMod, FastGreedy | 4 | Louvain, Oslom, **FastGreedy** |
| 8 | Copra | 7 | InfoMod |
| | | 8 | Copra |

Amongst the three measures we modified, the NMI appears to be the one leading to the results the most consistent with the observations previously made in [7]. Indeed, it seems to roughly preserve the order established by the traditional version, while distinguishing between otherwise not significantly different results, in a way compatible with our knowledge of the community structures topology. However, there is still room for improvement, since it is not able to separate Louvain, Oslom and FastGreedy. The two other measures are less satisfying, and display some anomalies. For instance, we cannot find an explanation for the very strong increase observed for Louvain, considering the topology of the communities it identified is relatively different from the reference.

## 6      Conclusion

In this article, we focused on the measures used to assess community detection algorithms. All those mentioned in the literature are similar to those used in data mining, more precisely in cluster analysis. Our first contribution is to have shown none of them is fully appropriate for this task, because they completely ignore network topology. This decreases their relevance, and can lead the user to incorrectly interpret the obtained scores. Our second contribution is to have defined variants of the three most widespread measures (F-Measure, Adjusted Rand index, Normalized mutual information), in order to solve this problem. For this matter, we modified them by introducing nodal weights: a different value can be associated to each node, allowing to penalize classification errors in an individual way. Adapting those modified measures to community detection is then straightforward: we need the weight to represent the topological importance of the node. We propose to use a combination of the degree and community embeddedness of the node. Our third contribution is the experimental evaluation of the proposed modifications. We used data obtained by applying a selection of community detection algorithms to a set of artificially generated networks with realistic topological properties. We compared the obtained rankings with those of the traditional versions of the measures, and assessed their consistency with observations from a previously conducted study regarding the topological properties of the estimated community structures [7]. On these data, the results obtained with the F-Measure and ARI present some inconsistencies. On the contrary, the modified version of the NMI is generally able to appropriately combine both aspects, i.e. assess how good the correspondence with the reference is in terms of both community membership and topological properties.

One of the limitations of this work concerns the size of the dataset used to evaluate our measures. To draw more definitive conclusions, it is necessary to test them on a larger corpus. Defining the weights used to introduce the topological aspect in the measures constitutes another sensitive point. Indeed, each weight is supposed to represent the importance of the associated node in the network, and this notion is difficult to define objectively. In this article, we penalized algorithms unable to treat correctly the nodes supposedly easy to qualify: those located at the core of the communities. But it would be possible to use the opposite approach, if we suppose all algorithms are able to correctly classify these nodes: one should then give more importance to those located on the border of the communities. We would then probably obtain rather different results.

Besides those points, our work can be extended in several ways. First, our modified measures can be used, as is, for different purposes. They were designed to compare an estimated and a reference partitions, but they could also be applied to two estimated partitions. One would then take the topological aspect into account when performing the comparison. The modified measures could also be used in the context of classic cluster analysis, i.e. on non-relational data, when one wants to distinguish the classified objects in terms of importance. Second, in this article we focused on plain networks, but the weights (and therefore the measures) could be adapted to various types of networks such as directed or weighted ones. Third, the principle of our modification could be applied to any other measure coming from cluster analysis. We only treated the most widespread in the community detection field, but many other exist: precision, recall, Jaccard index, etc.

**Acknowledgments.** This article is a translated and extended version of a previous work presented at the MARAMI 2012 conference [27].

## Appendix

This section contains material which was cut in the submitted version of this article. This mainly concerns the traditional and modified versions of the Rand index.

**Traditional and Modified (Non-Adjusted) Rand Indices**

Both the traditional and modified (non-adjusted) Rand indices were not presented in the experimental part, because they result in a lack of discrimination between the algorithms (as already observed in [7]). For this reason, they were cut from sections 2 and 4. For matters of completeness, here is the traditional version [5]:

$$RI(X,Y) = \left[\binom{n}{2} + 2\sum_{ij}\binom{n_{ij}}{2} - \sum_i \binom{n_{i+}}{2} - \sum_j \binom{n_{+j}}{2}\right] \Big/ \binom{n}{2} \qquad (29)$$

And here is the modified version, derived by replacing $W$ in (29) as explained in section 4.2:

$$RI'(X,Y) = \left[W(S) + 2\sum_{ij} W(x_i \cap y_j) - \sum_i W(x_i) - \sum_j W(y_j)\right] \Big/ W(S) \qquad (30)$$

The values obtained for the example of section 3 (as presented in Table 1 for the other measures) are respectively: 0.80 for the traditional version, 0.72 for the modified version applied to partition $A$ and 0.88 for the modified version applied to partition $B$.

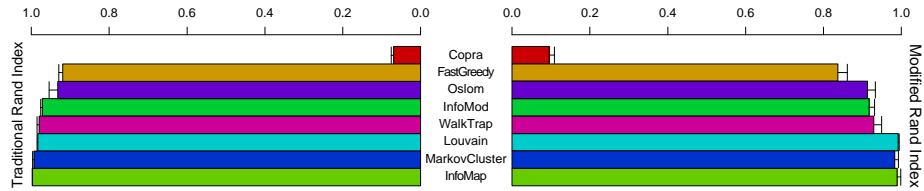

**Fig. 3.** Comparison of the results obtained with the traditional (left) and modified (right) versions of the (non-adjusted) Rand index.

The above figure represents the experimental results obtained for the traditional and modified versions of the (non-adjusted) Rand index, with the data presented in section 5. On the considered data, the measure does not seem to have a strong discriminant power. This is confirmed by Student's $t$ test, as displayed in Table 5: more than half the algorithms performances are not significantly different. The modified version of the measure distinguishes more groups, but we nevertheless decided not to include the (non-adjusted) Rand index in our study, and to focus only on its adjusted version instead.

**Table 5.** Algorithm rankings obtained with both traditional and modified versions of the (non-adjusted) Rand index. Algorithms experiencing a change in their relative position are represented in bold.

| Traditional Rand Index | | Modified Rand Index | |
|---|---|---|---|
| Rank | Algorithm | Rank | Algorithm |
| 1 | InfoMap, MarkovCluster, Louvain, WalkTrap, InfoMod | 1 | InfoMap, MarkovCluster, Louvain |
| 6 | Oslom, FastGreedy | 4 | **WalkTrap**, Oslom, **InfoMod** |
| 8 | Copra | 7 | **FastGreedy** |
|  |  | 8 | Copra |

**Comparison of Experimental Results**

The below table displays the rankings obtained with the four original and modified measures. It is a synthesis of the tables presented in section 5, plus the results obtained for the (non-adjusted) Rand index.

**Table 6.** Algorithm ranking obtained with the considered and proposed measures: F-Measure, Rand Index, Adjusted Rand Index, and Normalized Mutual Information.

| Algorithm | Traditional Versions | | | | Modified Versions | | | |
|---|---|---|---|---|---|---|---|---|
|  | FM | RI | ARI | NMI | FM | RI | ARI | NMI |
| Copra | 8 | 8 | 8 | 8 | 7 | 8 | 8 | 8 |
| FastGreedy | 6 | 6 | 5 | 6 | 6 | 7 | 6 | 4 |
| InfoMap | 1 | 1 | 1 | 1 | 1 | 1 | 1 | 1 |
| InfoMod | 6 | 1 | 5 | 6 | 7 | 4 | 7 | 7 |
| Louvain | 4 | 1 | 3 | 4 | 2 | 1 | 1 | 4 |
| MarkovCluster | 2 | 1 | 1 | 2 | 2 | 1 | 1 | 3 |
| Oslom | 4 | 6 | 5 | 4 | 4 | 4 | 4 | 4 |
| WalkTrap | 2 | 1 | 3 | 2 | 4 | 4 | 4 | 2 |